\documentstyle[preprint,tighten,eqsecnum,floats,epsfig,aps]{revtex}

\newcommand{\beq}{\begin{equation}}
\newcommand{\eeq}{\end{equation}}
\newcommand{\mc}{\multicolumn}
\def\dsla{\rlap{/}\partial}
\begin{document}
\draft

\title{Functional approach to the non-mesonic decay of 
${\bf \Lambda}$-hypernuclei}
\author{W. M. Alberico, A. De Pace, G. Garbarino}
\address{Dipartimento di Fisica Teorica, Universit\`a di Torino \\
and INFN, Sezione di Torino, 10125 Torino, ITALY}
\author{R. Cenni}
\address{Dipartimento di Fisica dell'Universit\`a di Genova \\
and INFN, sezione di Genova, 33-16146 Genova, ITALY}

\date{\today}
\maketitle

\begin{abstract}
We present an evaluation of the non-mesonic decay widths for $\Lambda$-hypernuclei
($\Lambda N\rightarrow NN$, $\Lambda NN\rightarrow NNN$) 
within the framework of the polarization propagator method.
The full $\Lambda$ self-energy
is evaluated microscopically in nuclear matter by using the functional approach, which 
supplies a theoretically well grounded approximation scheme for 
the classification of the
relevant diagrams, according to the prescriptions of the 
bosonic loop expansion. We employ average Fermi momenta, suitably adapted to
different mass number regions (medium-light, medium and heavy hypernuclei).
Moreover, we study the dependence of the decay 
rates on the $NN$ and ${\Lambda}N$ short range
correlations. With a proper choice of the parameters which control
these correlations in the new approximation scheme, 
it is possible to reproduce the experimental decay widths 
for $A\gtrsim 10$ hypernuclei.
\end{abstract}

\newpage
\pagestyle{plain}
\baselineskip 16pt
\vskip 48pt

\newpage
\section{Introduction}

Among the existing hypernuclei, those which contain 
one ${\Lambda}$ hyperon are the most stable with respect to the 
strong interaction and they are the subject of this paper.
The study of hypernuclear physics is connected to various and general
problems of both nuclear and particle physics. 
In fact, it helps in understanding important
aspects of weak interactions in nuclei, the role of sub-nucleonic 
degrees of freedom in the hadron-hadron interactions and
the renormalization properties of mesons and hyperons in the nuclear medium. 

The most interesting hypernuclear decays are those involving weak processes,
which directly concern the hyperon.
The subject of the weak decay rates of hyperons embedded in nuclei has a quite
long history, both on the theoretical and the experimental sides, but it has received
a broader attention only in the last ten years.  
The weak decay of $\Lambda$-hypernuclei occurs
through two different modes: the so called mesonic channel: 
\beq
\Lambda\rightarrow \pi N \hspace{0.1in} (\Gamma_M),
\eeq
and the non-mesonic one, which can be mainly attributed to the following processes:
\beq
\Lambda N \rightarrow NN \hspace{0.1in} (\Gamma_1),
\eeq
\beq
\Lambda NN \rightarrow NNN \hspace{0.1in} (\Gamma_2),
\eeq
both mediated by the exchange of $\pi, \rho, \eta, \omega, K, K^*$, etc. 
The non-mesonic mode is only possible in nuclear systems and,
nowadays, due to the difficulty of using $\Lambda$ beams, 
the study of the $\Lambda$ decay in nuclei is the only way to get information on the
weak process ${\Lambda}N \rightarrow NN$.

The free ${\Lambda}$ decay is
compatible with the ${\Delta}I=1/2$ isospin rule, which is based 
on the observed ratio 
$\Gamma_{\Lambda \rightarrow \pi^- p}/\Gamma_{\Lambda \rightarrow \pi^0 n}\simeq 2$,
but it is not yet understood on theoretical grounds. 
{}Besides, because of nuclear shell effects, 
it is not yet clear the level of violation of this rule for the mesonic decay
in nuclear systems.
In the present calculation we will assume this rule as valid. 
The momentum of the outgoing nucleon
in the process ${\Lambda}\rightarrow \pi N$ is about $100$ MeV,
hence this decay mode is suppressed by the Pauli principle in nuclei,
particularly in heavy systems (in infinite nuclear matter being strictly forbidden): 
the mesonic width is found to decrease fast as 
the mass number $A$ of the hypernucleus increases \cite{Al99,Os93,It95}. This is
confirmed by the few available experimental data.

In the non-mesonic process
${\Lambda}N\rightarrow NN$, the final nucleons have large
momenta ($\simeq 420$ MeV) and, apart from the $s$-shell hypernuclei, 
it dominates over the mesonic decay. 
The presence of large momentum transfers in the non-mesonic channels implies that 
the details of the nuclear structure
have only little influence on the decay, but, on the other hand,
the $NN$ and ${\Lambda}N$ short range correlations 
are crucial. There appears to be an
anticorrelation between mesonic and non-mesonic decay, such that the total width
is quite stable from light to heavy hypernuclei. In fact, as discussed in 
ref.~\cite{Al99}, the non-mesonic rates $\Gamma_1$ and $\Gamma_2$ increase and
reach saturation values as the mass number of the nucleus increases: 
$\Gamma_{NM}(^{208}_{\Lambda}{\rm Pb})/\Gamma_{NM}(^{12}_{\Lambda}{\rm C})
\simeq 1.4$, where $\Gamma_{NM}=\Gamma_1+\Gamma_2$.

In this paper we present an evaluation of the decay rates for a $\Lambda$ in nuclear
matter within the framework of the polarization propagators. 
The relevant Feynman diagrams have been obtained following a functional approach, 
which allows to divide these diagrams into classes, according to the
prescription of the so-called bosonic loop expansion (BLE). 
The calculation in finite
nuclei (using the local density approximation) 
is not possible here because of the long computing time already for the calculation
of the decay widths at fixed Fermi momentum (namely in nuclear matter).
Nevertheless, in order to consider
different mass regions, we assign an {\sl average} Fermi momentum 
felt by the hyperon in the various nuclei.

We remind that the baryon-baryon strong interactions cannot be treated with the 
standard perturbative method. However, in the study of nuclear phenomena 
we always need to sum up the relevant diagrams. For instance,one usually performs
the summation of the infinite classes of diagrams entailed by the RPA and
Hartree-Fock approximations. However, in the above quoted schemes no prescription is given
to evaluate the ``next-to-leading'' order. 

The functional techniques can provide a theoretically founded derivation of 
new classes of expansion in terms of powers of suitably chosen parameters. 
On the other hand, as we will see, the RPA
automatically appears in this framework as the mean field level. 
The method has been extensively applied
to different processes in nuclear physics \cite{Ne82,Al87,Ce97}. 
We will use it for the calculation of the $\Lambda$ decay rates in nuclear
matter, which can be expressed through the nuclear responses 
to pseudoscalar-isovector and vector-isovector fields. 
We will see that the responses include ring-dressed meson propagators 
(which represent the mean field level of the theory)
and almost the whole spectrum of {\sl 2p-2h} excitations (expressed in terms of a 
one-loop expansion with respect to the ring-dressed meson propagators)
which are required for the evaluation of $\Gamma_2$. 
Actually, the semiclassical expansion
leads to the prescription of grouping the relevant Feynmam diagrams in a consistent 
many-body description of the ``in medium'' meson self-energies: 
the general theorems and sum rules of the theory are preserved.

The paper is organized as follows. In Sec.~\ref{pm} we summarize
the model used for the calculation of the decay rates. 
In Sec.~\ref{func} the functional approach to the spin-isospin nuclear
response functions in presented. By using the semiclassical approximation we
will introduce the class of Feynman diagrams needed for the evaluation 
of $\Gamma_2$.
Our results are presented and discussed in Sec.~\ref{res}: we first study
the sensitivity of the decay rates to the $NN$
and $\Lambda N$ short range correlations, and then we parametrize these
correlations in order to reproduce 
the experimental decay widths for three different mass regions in the hypernuclear
spectrum. Our conclusions are given in Sec.~\ref{concl}. 

\section{Polarization propagator method}
\label{pm}
In this section we briefly summarize how the $\Lambda$ decay in nuclear
matter can be studied employing the polarization propagator method \cite{Os85,Ra95,Al99}.
This technique provides a unified picture
of the different decay channels and it is equivalent to the standard
wave function method, used by other authors in 
refs.~\cite{Os93,Pa98,Pa97,Du96,It95}. In a previous work \cite{Al99}
the evaluation of the decay widths was first performed in nuclear matter and
then extended to finite nuclei via the local density approximation:
in this framework the experimental partial decay rates in a range of nuclei 
from $^{12}_{\Lambda}$C to
$^{208}_{\Lambda}$Pb have been reproduced.
The two-body induced decay $\Lambda NN\rightarrow NNN$
has been accounted for via a purely phenomenological parametrization 
(using data on pionic atoms) of the
{\sl 2p-2h} polarization propagator which enters
the $\Lambda$ self-energy $\Sigma_{\Lambda}$. The total
decay rate is obtained through the relation:
\beq
\label{Gamma}
{\Gamma}_{\Lambda}=-2{\rm Im}{\Sigma}_{\Lambda} .
\eeq
Here we present 
a microscopic calculation of the {\sl 2p-2h} polarization propagator 
within the theoretically
consistent scheme which will be introduced in the next section.

\begin{figure}
\begin{center}
\mbox{\epsfig{file=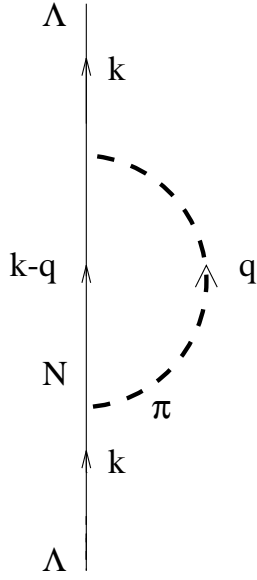,width=.20\textwidth}}
\vskip 2mm
\caption{$\Lambda$ self energy in nuclear matter (the dashed line represents
a dressed pion).}
\label{self}
\end{center}
\end{figure}

\begin{figure}
\begin{center}
\mbox{\epsfig{file=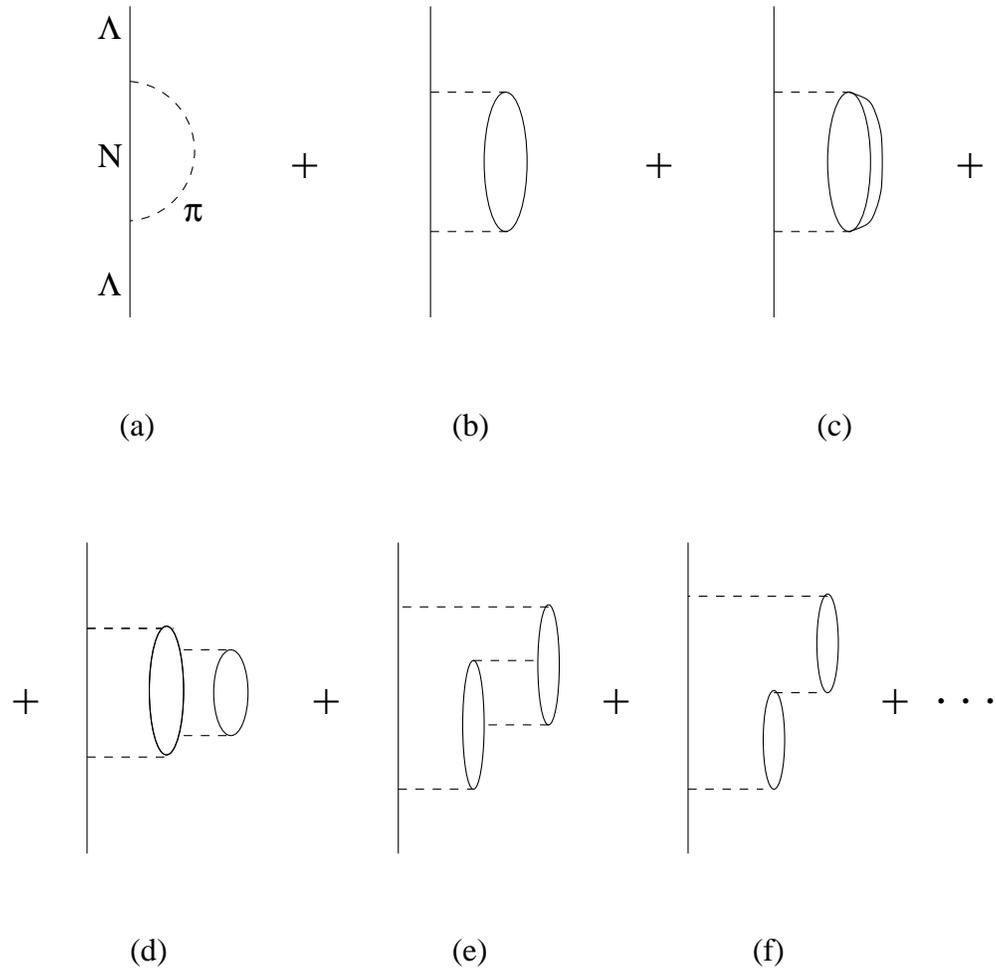,width=.8\textwidth}}
\vskip 6mm
\caption{Lowest order terms for the ${\Lambda}$ self-energy in 
nuclear matter. The meaning of the various diagramms is explained in the text.}
\label{self11}
\end{center}
\end{figure}

Let us first remind the main steps in the evaluation of $\Gamma_{\Lambda}$.
We start from the ${\Lambda}\rightarrow \pi N$ effective lagrangian:
\beq
\label{lagran}
{\cal L}_{{\Lambda}\pi N}=G m_{\pi}^2\overline{\psi}_N(A+B\gamma_5)
{\bbox \tau} \cdot {\bbox \phi}_{\pi}{\psi}_{\Lambda}+h.c. ,
\eeq
where the values of the weak coupling constants 
$G\simeq 2.211\cdot 10^{-7}/m_{\pi}^2$, $A=1.06$, 
$B=-7.10$ are fixed from the free ${\Lambda}$ decay. The constants $A$ and $B$ 
determine the strengths of the parity
violating and parity conserving ${\Lambda}\rightarrow \pi N$ amplitudes,
respectively.
In order to enforce the ${\Delta}I=1/2$ rule, in eq.~(\ref{lagran}) 
the hyperon is assumed to be an isospin spurion with $I_z=-1/2$.
Using the Feynman rules, in the non relativistic limit
the ${\Lambda}$ self-energy in nuclear matter is (see fig.~\ref{self}):
\beq
\label{Sigma1}
{\Sigma}_{\Lambda}(k)=3i(G m_{\pi}^2)^2\int \frac{d^4q}{(2\pi)^4}
\left\{S^2+\frac{P^2}{m_{\pi}^2}\bbox q^2\right\}F_{\pi}^2(q)
G_N(k-q)G_{\pi}(q) .
\eeq
Here, $S=A$, $P=m_{\pi}B/2m_N$, while:
\beq
\label{propnucl}
G_N(p)=\frac{{\theta}(\mid \bbox p \mid-k_F)}{p_0-E_N(\bbox p)-V_N+i{\epsilon}}+
\frac{{\theta}(k_F-\mid \bbox p \mid)}{p_0-E_N(\bbox p)-V_N-i{\epsilon}} ,
\eeq
and:
\beq
\label{proppion}
G_{\pi}(q)=\frac{1}{q_0^2-\bbox q^2-m_{\pi}^2-{\Sigma}_{\pi}^*(q)} ,
\eeq
are the nucleon and pion propagators in nuclear matter, respectively.
In the above, $p=(p_0,\bbox p)$ and $q=(q_0,\bbox q)$ denote 
four-vectors, $k_F$ is the Fermi momentum, $E_N$ is the nucleon total free
energy, $V_N$ is the nucleon binding energy (assumed constant), 
and ${\Sigma}_{\pi}^*$ is the pion proper self-energy. 
Moreover, in eq.~(\ref{Sigma1}) we have included a monopole form factor for 
the $\pi\Lambda N $ vertex:
\beq
\label{ff}
F_{\pi}(q)=\frac {{\Lambda}_{\pi}^2-m_{\pi}^2}{{\Lambda}_{\pi}^2-q_0^2+\bbox q^2} ,
\eeq
with the same cut-off ${\Lambda}_{\pi}=1.3$ GeV used for the $\pi NN$ strong vertex.
In fig.~\ref{self11} we explicitely show
the lowest order Feynman graphs for the ${\Lambda}$ 
self-energy. Diagram (a) represents the bare
self-energy term, including the effects of Pauli principle and of
the binding on the intermediate nucleon.
In (b) and (c) the pion couples to a particle-hole ({\sl p-h}) and a 
{\sl ${\Delta}$-h} pairs, respectively. In diagrams (d) and (e) we 
show examples of {\sl 2p-2h} 
excitations, while (f) is a RPA iteration of diagram (b). 
Once evaluated the integral over $q_0$ in (\ref{Sigma1}), the nuclear matter 
${\Lambda}$ decay width (eq.~(\ref{Gamma})) becomes \cite{Os85}:
\begin{eqnarray}
\label{Sigma2}
{\Gamma}_{\Lambda}(\bbox k,\rho)&=&-6(G m_{\pi}^2)^2\int \frac{d\bbox q}
{(2\pi)^3}{\theta}(|\bbox k- \bbox q| -k_F)
{\theta}(k_0-E_N(\bbox k-\bbox q)-V_N) \nonumber
\\
& & \times {\rm Im}{\alpha}(q)\mid _{q_0=k_0-E_N(\bbox k-\bbox q)-V_N} ,
\end{eqnarray}
where
\begin{eqnarray}
\label{Alpha}
{\alpha}(q)&=&\left\{S^2+\frac{P^2}{m_{\pi}^2}\bbox q^2\right\}F_{\pi}^2(q)
G_{\pi}^0(q)+\frac{\tilde{S}^2(q)U_L(q)}{1-V_L(q)U_L(q)} \nonumber \\
& & +\frac{\tilde{P}_L^2(q)U_L(q)}{1-V_L(q)U_L(q)}+
\frac{\tilde{P}_T^2(q)U_T(q)}{1-V_T(q)U_T(q)} ,
\end{eqnarray}
consist of a longitudinal and a transverse part.
In eq.~(\ref{Sigma2}) the first ${\theta}$-function forbids
intermediate nucleon momenta (see fig.~\ref{self}) smaller than the 
Fermi momentum and the second one requires the pion energy $q_0$ to be 
positive. Moreover, the ${\Lambda}$ energy, 
$k_0=E_{\Lambda}(\bbox k)+V_{\Lambda}$,
contains a binding term. The pion lines of fig.~\ref{self11} 
have been replaced in eq.~(\ref{Alpha}) by the interactions 
$V_L$, $V_T$, $\tilde{S}$, $\tilde{P}_L$, $\tilde{P}_T$,
which include $\pi$ and $\rho$ exchange modulated by the effect of short range
correlations. They have the following expressions:
\beq
\label{vl}
V_L(q)=\frac{f_{\pi}^2}{m_{\pi}^2}\left\{\bbox q^2F_{\pi}^2(q)G_{\pi}^0(q)+
g_L(q)\right\} ,
\eeq
\beq
\label{vt}
V_T(q)=\frac{f_{\pi}^2}{m_{\pi}^2}\left\{\bbox q^2C_{\rho}F_{\rho}^2(q)
G_{\rho}^0(q)+g_T(q)\right\} .
\eeq
\beq
\label{pl}
\tilde{P}_L(q)=\frac{f_{\pi}}{m_{\pi}}\frac{P}{m_{\pi}}
\left\{\bbox q^2F_{\pi}^2(q)G_{\pi}^0(q)+g_L^{\Lambda}(q)\right\} ,
\eeq
\beq
\label{pt}
\tilde{P}_T(q)=\frac{f_{\pi}}{m_{\pi}}\frac{P}{m_{\pi}}g_T^{\Lambda}(q) ,
\eeq
\beq
\label{s}
\tilde{S}(q)=\frac{f_{\pi}}{m_{\pi}}S\left\{F_{\pi}^2(q)G_{\pi}^0(q)-
\tilde{F}_{\pi}^2(q)\tilde{G}_{\pi}^0(q)\right\}\mid \bbox q \mid .
\eeq
In the above, $G_{M}^0=1/(q_0^2-\bbox q^2-m_{M}^2)$ denotes a
free meson propagator and $F_{\rho}$ is the $\rho NN$ form factor (eq.~(\ref{ff})
with cut-off ${\Lambda}_{\rho}=2.5$ GeV). The exchange of $\rho$-mesons is brought
into play by the short range correlation $\tilde P_T$ embodied in the 
$\Lambda N\rightarrow NN$ potential.
Form factors and propagators with a tilde imply that they are calculated
with the replacement 
$\bbox q^2\rightarrow~\bbox q^2+q_c^2$, where $1/q_c$ is a lenght
of the order of the hard core radius of the interaction, and
$C_{\rho}$ is the ratio:
\beq
\label{rhocoup}
C_{\rho}={\frac{f_{\rho}^2}{m_{\rho}^2}}
\left[\frac{f_{\pi}^2}{m_{\pi}^2}\right]^{-1} .
\eeq

The short range correlations included in the potentials of 
eqs.~(\ref{vl})-(\ref{pt}) are explicitely reported in the appendix 
of ref.~\cite{Al99}. Their momentum dependence (governed by the
parameter $q_c$) ensures the convergence of the integrals contained in 
diagrams like (d) and (e) of fig.~\ref{self11} (this behaviour is not crucial
in ring approximation). In the calculation, the
values of the correlation functions in $q=0$,
\beq
g^{\prime}\equiv g_{L,T}(0), \hspace{0.2 in} g^{\prime}_{\Lambda}\equiv 
g^{\Lambda}_{L,T}(0)
\eeq
(namely the Landau parameters), will be considered as free. 

Furthermore, in eq.~(\ref{Alpha}) $U_L$ and $U_T$ contain the Lindhard 
functions for {\sl p-h} and {\sl ${\Delta}$-h} 
excitations \cite{Wa71} and also account for the irriducible {\sl 2p-2h} 
polarization propagator:
\beq
\label{propU}
U_{L,T}(q)=U^{ph}(q)+U^{\Delta h}(q)+U_{L,T}^{2p2h}(q) .
\eeq
They appear in eq.~(\ref{Alpha}) within the standard RPA expression.
The Lindhard functions $U^{ph}$, $U^{\Delta h}$ are normalized as
in ref.~\cite{Os90}.
In the following section we will develop an approximation scheme for the
classification of the various diagrams to be included in the calculation of 
the $\Lambda$ self-energy.
Eq.~(\ref{Sigma2}) depends, both explicitly and through $U_{L,T}(q)$,
on the nuclear matter density $\rho=2k_F^3/3{\pi}^2$.

We remind that in nuclear matter the mesonic decay is strictly forbidden,
the final nucleon in $\Lambda\rightarrow \pi N$ having a momentum
lower than the Fermi one (for a $\Lambda$ decaying at rest, 
$q\simeq 100$ MeV $< k_F\simeq 270$ MeV).
In finite nuclei the mesonic rate is dominant only in $s$-shell hypernuclei and
rapidly decreases with the mass number. 

\section{Functional approach to the spin-isospin nuclear response functions}
\label{func}

We first evaluate the polarization propagator in the
pionic (spin-longitudinal) channel. 
In order to exemplify we consider a lagrangian describing a system of 
nucleons interacting with pions through a pseudoscalar
coupling:
\beq
{\cal L}_{\pi N}=\overline \psi (i\dsla-m_N)\psi 
+\frac{1}{2}\partial_{\mu}{\bbox \phi}\cdot{\partial^{\mu} {\bbox \phi}}
-\frac{1}{2}m_{\pi}^2 \bbox\phi^2-i\overline \psi \bbox\Gamma\psi \cdot \bbox \phi , 
\eeq
where $\psi$ ($\bbox \phi$) is the nucleonic (pionic) field, and:
\beq
\bbox\Gamma=g\gamma_5\bbox\tau 
\eeq
($g=2f_{\pi}m_N/m_{\pi}$)
is the spin-isospin matrix in the spin-longitudinal isovector channel. 
We remind that
in the calculation of the hypernuclear decay rates we also 
need the polarization propagator in the transverse channel 
(see eqs.~(\ref{Sigma2}),(\ref{Alpha})): hence, we will have to include in the
model another mesonic degree of freedom, 
the $\rho$ meson. As we will see, the semiclassical expansion 
is characterized by the
topology of the diagrams, so the same scheme can be applied to mesonic fields
other than the pionic one.
In this section we use a relativistic formalism, the non-relativistic
reduction of our results being trivial. 

Let us now introduce a {\it classical} external field $\bbox\varphi$ with the quantum numbers
of the pion. The lagrangian then becomes:
\beq
{\cal L}_{\pi N} \rightarrow {\cal L}_{\pi N}-i\overline \psi 
\bbox\Gamma\psi \cdot \bbox \varphi.
\eeq
The corresponding generating functional in terms of Feynman path integrals has the form:
\beq
\label{Z}
Z[\bbox \varphi]=\int {\cal D} \left[\overline \psi, \psi, \bbox \phi\right]
\exp\left\{i \int dx \left[{\cal L}_{\pi N}(x)-i
\overline \psi(x) \bbox\Gamma\psi(x) \cdot \bbox \varphi(x)\right]\right\} 
\eeq
(here and in the following the coordinate integrals are 4-dimensional).
All the fields in the functional integrals have to be considered as classical
variables, but with the correct commuting properties. The physical quantities
of interest for the problem are deduced from the generating functional by means
of functional differentiations. By introducing a new functional $Z_c$ such that
\beq
\label{connect}
Z[\bbox\varphi]=\exp{\left\{iZ_c[\bbox\varphi]\right\}} ,
\eeq
the polarization propagator turn out to be the second 
functional derivative of $Z_c$ with respect to the source $\bbox\varphi$ of the
pionic field:
\beq
\label{proppol}
\Pi_{ij}(x,y)=-\left[\displaystyle \frac{\delta^2 Z_c[\bbox\varphi]}{\delta \varphi_i(x) 
\delta \varphi_j(y)}\right]_{\bbox \varphi=0}  .
\eeq

We notice that the use of $Z_c$ instead of $Z$ in eq.~(\ref{proppol}) amounts to cancel the
disconnected diagrams of the perturbative expansion (linked cluster theorem). 
From the generating functional $Z$ one can obtain different 
approximation schemes according 
to the order in which the functional integrations are performed. By
integrating eq.~(\ref{Z}) over the mesonic degrees of freedom {\it first}, 
the generating functional
can be written in terms of a fermionic effective action $S^F_{\rm eff}$. 
Up to an irrelevant multiplicative 
constant:
\beq
Z[\bbox\varphi]=\int {\cal D} \left[\overline \psi, \psi\right]
\exp\left\{i S^F_{\rm eff}\left[\overline \psi, \psi\right]\right\} .
\eeq
The remaining integration variables are interpreted as 
physical fields and $S^F_{\rm eff}$ describes a quadrilinear non-local time- or 
energy-dependent
nucleon-nucleon interaction induced by the exchange of one pion:
\begin{eqnarray}
\label{SeffF}
S^F_{\rm eff}\left[\overline \psi, \psi \right]&=&\int dx\,dy\, 
\left[\overline\psi(x) \right. G^{-1}_N(x-y)\psi(y) \nonumber \\
& & +\frac{1}{2}\sum_{i=1}^{3}\overline \psi(x) \Gamma_i\psi(x) G^0_{\pi}(x-y)
\left.\overline \psi(y) \Gamma_i\psi(y)\right] ,
\end{eqnarray}
where $G_N$ and $G^0_{\pi}$ are the nucleon and free pion propagators, 
respectively, which satisfy the following field equations: 
\beq
\left(i\dsla_x-m_N-i
\bbox\Gamma \cdot \bbox \varphi\right)G_N(x-y)=\delta(x-y) ,
\eeq
\beq
\left(\Box_x+m^2_{\pi}\right)G^0_{\pi}(x-y)=-\delta(x-y) .
\eeq
The pion propagator is diagonal in the isospin
indices: $\left(G^0_{\pi}\right)_{ij}=\delta_{ij}G^0_{\pi}$.
The effective action (\ref{SeffF}) can then be utilized in the framework
of ordinary perturbation theory  and will not be employed in the following.
In fact, it does not bring significant novelties with respect to the usual 
calculations;
furthermore, it cannot be correctly renormalized due to the absence of a
term proportional to $\bbox\phi^4$, which is needed 
to cancel the divergence of the 4-points
fermion loops.

\subsection{The bosonic effective action}

\begin{figure}
\begin{center}
\mbox{\epsfig{file=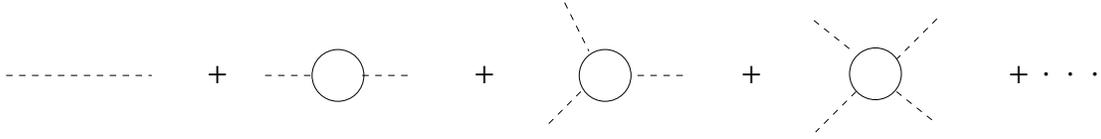,width=.9\textwidth}}
\vskip 2mm
\caption{Diagrammatic representation of the bosonic effective action 
\protect(\ref{seffb}).}
\label{vpi}
\end{center}
\end{figure}

Alternatively, it is possible to eliminate in eq.~(\ref{Z})
the nucleonic degrees of freedom first (without destroying 
the renormalizability of the theory\cite{Al87}). Introducing the change of
variable $\bbox\phi\rightarrow \bbox\phi-\bbox\varphi$, eq.~(\ref{Z}) becomes:

\begin{eqnarray}
\label{Z1}
Z[\bbox \varphi]&=&\exp\left\{\frac{i}{2}\int dx\,dy\, \bbox\varphi(x) \cdot
G^{0^{-1}}_{\pi}(x-y)\bbox\varphi(y)\right\} \nonumber \\ 
& & \times \int {\cal D} \left[\overline \psi, \psi, \bbox \phi\right]
\exp\left\{i \int dx\,dy\, \right.\left[\overline\psi(x)G^{-1}_N(x-y)\psi(y)\right. 
\nonumber \\
& & \left. \left. +\frac{1}{2}\bbox\phi(x)\cdot G^{0^{-1}}_{\pi}(x-y)
\left(\bbox\phi(y)+2\bbox\varphi(y)\right)\right]\right\} ,
\end{eqnarray}
where the integral over $\left[\overline\psi, \psi\right]$ is gaussian:

\beq
\int {\cal D} \left[\overline \psi, \psi \right] 
\exp\left\{i \int dx\,dy\, \overline\psi(x)G^{-1}_N(x-y)\psi(y)\right\}=
\left({\rm det} G_N \right)^{-1} .
\eeq
Hence, multiplying eq.~(\ref{Z1}) by the unessential factor 
det~$G^0_N$ ($G^0_N$ being the free nucleon propagator), 
which only redefines the normalization constant of the generating functional, and using the
property ${\rm det} X=\exp\left\{{\rm Tr}\ln X\right\}$, we obtain:

\beq
\label{Z2}
Z[\bbox \varphi]=\exp\left\{\frac{i}{2}\int dx\,dy\, \bbox\varphi(x)\cdot
G^{0^{-1}}_{\pi}(x-y)\bbox\varphi(y)\right\}
\int {\cal D} \left[\bbox \phi\right]
\exp\left\{i S^B_{\rm eff}\left[\bbox\phi \right]\right\} ,
\eeq
with:

\beq
\label{seffb}
S^B_{\rm eff}\left[\bbox\phi\right]=\int dx\,dy\, 
\left\{\frac{1}{2}\bbox\phi(x)\cdot G^{0^{-1}}_{\pi}(x-y)
\left[\bbox\phi(y)+2\bbox\varphi(y)\right]+V_{\pi}[\bbox\phi]\right\} , 
\eeq

\begin{eqnarray}
\label{pionpot}
V_{\pi}[\bbox\phi]&=&i{\rm Tr}\sum_{n=1}^{\infty}\frac{1}{n}
\left(i\bbox\Gamma\cdot\bbox\phi G^0_N\right)^n \nonumber \\
& =&\frac{1}{2}\sum_{i,j}{\rm Tr}\left(\Gamma_i\Gamma_j\right)\int dx\,dy\, \Pi^0(x,y)
\phi_i(x)\phi_j(y) \nonumber \\
& & +\frac{1}{3}\sum_{i,j,k}{\rm Tr}\left(\Gamma_i\Gamma_j\Gamma_k\right)
\int dx\,dy\,dz\, \Pi^0(x,y,z)\phi_i(x)\phi_j(y)\phi_k(z)+ 
{\cal O}(\bbox\phi^4) ,
\end{eqnarray}
where
\begin{eqnarray}
\label{pizero}
-i\Pi^0(x,y)&=&iG^0_N(x-y)iG^0_N(y-x) , \\
-i\Pi^0(x,y,z)&=&iG^0_N(x-y)iG^0_N(y-z)iG^0_N(z-x) ,\hspace{0.15 in} {\rm etc.}
\end{eqnarray}
We have thus derived an effective action for the bosonic field 
$\bbox\phi$. This action contains a term for the
free pion field and also a highly non-local pion self-interaction 
$V_{\pi}$, which is described by the
Feynman diagrams shown in fig.~\ref{vpi}. This effective interaction 
is given by the sum of all the diagrams containing one closed fermion loop.
We note that the function of eq.~(\ref{pizero}) is the free polarization
propagator, namely the Lindhard function. Moreover, the functions 
$\Pi^0(x,y,..,z)$ are symmetric for cyclic permutations of the arguments.
\subsection{Semiclassical expansion}

The next step is the evaluation of the functional integral over the
bosonic degrees of freedom in eq.~(\ref{Z2}). A perturbative approach
to the bosonic effective action (\ref{seffb}) does not seem to provide any
valuable results within the capabilities of the present computing tools
and we will use here another approximation scheme, namely the
semiclassical method. 
\subsection*{Mean field level}
The lowest order of the semiclassical expansion is the stationary phase
approximation (also called saddle point approximation (SPA) in the euclidean space): 
we require the bosonic effective action to be stationary with respect to
arbitrary variations of the fields $\phi_i$:

\beq
\displaystyle\frac{\delta S^B_{\rm eff}\left[\bbox\phi\right]}{\delta \phi_i(x)}=0 .
\eeq
From the partial derivative of eq.~(\ref{seffb}) we obtain for the classical 
field $\bbox\phi$
the following equation of motion:
\beq
\label{eqmoto}
\left(\Box+m^2_{\pi}\right)\phi_i(x)=\int dy\, G^{0^{-1}}_{\pi}(x-y)\varphi_i(y)+
\displaystyle \frac{\delta V_{\pi}\left[\bbox\phi\right]}{\delta \phi_i(x)} ,
\eeq
whose solutions are functional of the external source $\bbox\varphi$. The exact 
solution cannot be written down explicitly. However, due to the
particular form of $V_{\pi}[\bbox\phi]$, when $\bbox\varphi\rightarrow 0$,
one solution is $\bbox\phi=0$; we can then express the general 
solution of eq.~(\ref{eqmoto}) 
as an expansion in powers of $\bbox\varphi$:
\beq
\label{soluz}
\phi^0_i(x)=\sum_{j}\int dy\, A_{ij}(x,y)\varphi_j(y)+
\frac{1}{2}\sum_{j,k}\int dy\,dz\, B_{ijk}(x,y,z)\varphi_j(y)\varphi_k(z)+
{\cal O}\left(\bbox\varphi^3\right) .
\eeq
Substituting eqs.~(\ref{soluz}) and (\ref{pionpot}) 
into (\ref{eqmoto}) and keeping the terms linear in
$\varphi_i$ we obtain the following relation for $A_{ij}$:
\beq
\label{A}
A_{ij}(x,y)-{\rm Tr}\left(\Gamma_i^2\right)\int du\,dv\, 
G^0_{\pi}(x-u)\Pi^0(u,v)A_{ij}(v,y)=\delta_{ij}\delta(x-y) .
\eeq
Finally, by introducing the ring-dressed pion propagator 
$G^{\rm ring}_{\pi}$, which satisfies the Dyson equation:
\beq
G^{\rm ring}_{\pi}(x-y)=G^0_{\pi}(x-y)+{\rm Tr}\left(\Gamma_i^2\right)
\int du\,dv\, G^0_{\pi}(x-u)
\Pi^0(u,v)G^{\rm ring}_{\pi}(v-y) ,
\eeq
or, formally:
\beq
G^{\rm ring}_{\pi}=\displaystyle \frac{G^0_{\pi}}
{1-{\rm Tr}\left(\Gamma_i^2\right)G^0_{\pi}\Pi^0} \; ,
\eeq
the solution of (\ref{A}) reads:

\beq
\label{AA}
A_{ij}(x,y)=\delta_{ij}\int dz\, G^{\rm ring}_{\pi}(x-z) G^{0^{-1}}_{\pi}(z-y) .
\eeq
Therefore, the saddle point of (\ref{seffb}) at first order
in the source $\bbox\varphi$ is:

\beq
\label{soluzring}
\phi^{\rm ring}_i(x)=\int dy\,dz\, G^{\rm ring}_{\pi}(x-z) 
G^{0^{-1}}_{\pi}(z-y)\varphi_i(y)\equiv
\int dy\, \left(G^{\rm ring}_{\pi} G^{0^{-1}}_{\pi}\right)(x-y)\varphi_i(y) ,
\eeq
and the corresponding bosonic effective action reads:

\beq
S^B_{\rm eff}\left[\bbox\phi^{\rm ring}\right]=-\frac{1}{2}\int dx\,dy\,du\,dv\, 
G^{0^{-1}}_{\pi}(x-u)\bbox\varphi(u)\cdot G^{\rm ring}_{\pi}(x-y)
G^{0^{-1}}_{\pi}(y-v)\bbox\varphi(v) .
\eeq
Now, the generating functional of eq.~(\ref{Z2}) takes the form:

\begin{eqnarray}
Z\left[\bbox\varphi\right]&=&\exp\left\{\frac{i}{2}\right. \int dx\,dy\,du\,dv\,
\bbox\varphi(u)\cdot G^{0^{-1}}_{\pi}(x-u) \nonumber \\
& & \times \left. \left[G^0_{\pi}(x-y)-G^{\rm ring}_{\pi}(x-y)\right]
G^{0^{-1}}_{\pi}(y-v)\bbox\varphi(v)\right\} .
\end{eqnarray}
The polarization propagator is then evaluated using eqs.~(\ref{connect}), 
(\ref{proppol}).
We obtain that in the saddle point approximation it coincides with the well known
ring expression:
\begin{eqnarray}
\Pi_{ij}(x,y)&=&\delta_{ij}\left[\Pi^0(x,y)+{\rm Tr}\left(\Gamma^2_i\right)
\int du\,dv\, \Pi^0(x,u)G^{\rm ring}_{\pi}(u-v)\Pi^0(v,y)\right] \nonumber \\
& \equiv & \delta_{ij}\Pi^{\rm ring}(x,y) ,
\end{eqnarray}
or, formally:
\beq
\Pi=\displaystyle \frac{\Pi^0}
{1-{\rm Tr}\left(\Gamma_i^2\right)G^0_{\pi}\Pi^0}
\equiv \Pi^{\rm ring} .
\eeq
Hence, the ring approximation corresponds to the mean field level
of the present effective theory.
\subsection*{Quantum fluctuations around the mean field solution
(one-loop corrections)}

In the next step of our semiclassical expansion we write the bosonic 
effective action as:

\beq
S^B_{\rm eff}\left[\bbox\phi\right]=S^B_{\rm eff}\left[\bbox\phi^0\right]+
\frac{1}{2} \sum_{ij} \int dx\,dy\, \displaystyle 
\left[\frac{\delta^2S^B_{\rm eff}\left[\bbox\phi\right]}{\delta\phi_i(x)\delta\phi_j(y)}
\right]_{\bbox\phi=\bbox\phi^0} \left[\phi_i(x)-\phi^0_i(x)\right]
\left[\phi_j(y)-\phi^0_j(y)\right],
\eeq
where now $\bbox\phi^0$ also contains the second order term in the source 
$\bbox\varphi$ (see eq~(\ref{soluz})). Then, after performing the
gaussian integration over
$\bbox\phi$, the generating functional reads:

\begin{eqnarray}
\label{Z3}
Z\left[\bbox\varphi\right]&=&\exp\left\{\frac{i}{2}\int dx\,dy\, 
\bbox\varphi(x)\cdot G^{0^{-1}}_{\pi}(x-y)\bbox\varphi(y)\right\} \nonumber \\
& & \times \exp\left\{iS^B_{\rm eff}\left[\bbox\phi^0\right]
-\frac{1}{2}{\rm Tr}\ln \left[\displaystyle
\frac{\delta^2S^B_{\rm eff}\left[\bbox\phi\right]}
{\delta\phi_i(x)\delta\phi_j(y)}\right]_{\bbox\phi=\bbox\phi^0}\right\} ,
\end{eqnarray}
and the polarization propagator is:

\beq
\label{proppol1}
\Pi_{ij}(x,y)=-\left\{\frac{\delta^2}
{\delta \varphi_i(x)\delta \varphi_j(y)}\left[S^B_{\rm eff}
\left[\bbox\phi^0\right]+\frac{i}{2}{\rm Tr}\ln 
\left(\displaystyle \frac{\delta^2S^B_{\rm eff}\left[\bbox\phi\right]}
{\delta\phi_k(x)\delta\phi_l(y)}\right)_{\bbox\phi=\bbox\phi^0}\right]
\right\}_{\bbox\varphi=0} ,
\eeq
In the above the second derivative of the effective action (\ref{seffb}) at the
order $\bbox\phi^2$ turns out to be:

\begin{eqnarray}
\label{deriv2}
\frac{\delta^2S^B_{\rm eff}\left[\bbox\phi\right]}
{\delta\phi_i(x)\delta\phi_j(y)}&=&\delta_{ij}G^{0^{-1}}_{\pi}(x-y)
+{\rm Tr}\left(\Gamma_i\Gamma_j\right)\Pi^0(x,y) \nonumber \\
& &+\sum_{k} \int du \left[{\rm Tr}\left(\Gamma_i\Gamma_j\Gamma_k\right)\Pi^0(x,y,u)
+{\rm Tr}\left(\Gamma_j\Gamma_i\Gamma_k\right)\Pi^0(y,x,u)\right]\phi_k(u) \nonumber \\
& &+\sum_{k,l}\int du\,dv\, \left[{\rm Tr}\left(\Gamma_i\Gamma_j\Gamma_k\Gamma_l\right)
\Pi^0(x,y,u,v) \right.
+{\rm Tr}\left(\Gamma_j\Gamma_i\Gamma_k\Gamma_l\right)\Pi^0(y,x,u,v) \nonumber \\
& &+\left. {\rm Tr}\left(\Gamma_i\Gamma_l\Gamma_j\Gamma_k\right)\Pi^0(x,v,y,u)\right]
\phi_k(u)\phi_l(v) .
\end{eqnarray}
The second term in the r.h.s.~of eq.~(\ref{deriv2}) does not affect the calculation 
of eq.~(\ref{proppol1}). By substituting (\ref{soluz}) in
the equation of motion (\ref{eqmoto}), from the terms of order $\bbox\varphi^2$
we get the following $B_{ijk}$ functions:

\begin{eqnarray}
\label{BB}
B_{ijk}(x,y,z)&=&2{\rm Tr}\left(\Gamma_i\Gamma_j\Gamma_k\right)\int du\,dv\,dt\,
\Pi^0(u,v,t) \nonumber \\
& & \times G^{\rm ring}_{\pi}(x-u)\left(G^{\rm ring}_{\pi} 
G^{0^{-1}}_{\pi}\right)(v-y)
\left(G^{\rm ring}_{\pi} G^{0^{-1}}_{\pi}\right)(t-z) .
\end{eqnarray}
Now we have to calculate the logarithm in eq.~(\ref{proppol1}) up to second 
order in $\bbox\varphi$. We can multiply the generating functional (\ref{Z3}) 
by the factor $\left({\rm det} G^0_{\pi}\right)^{-1/2}$, inessential in the 
calculation of the polarization propagator (this corresponds to multiply
eq.~(\ref{deriv2}) by $G^0_{\pi}$). Then, after calculating eq.~(\ref{deriv2})
for $\bbox\phi=\bbox\phi^0$, with  $\bbox\phi^0$ given by the 
eqs.~(\ref{soluz}), (\ref{AA}), (\ref{BB}),
we expand the logarithm up to $\bbox\varphi^2$ and take the trace 
to the same order. Finally, the derivation with respect to the external
source provides the following total polarization propagator:
\beq
\Pi_{ij}(x,y)=\delta_{ij}\Pi (x,y),
\eeq
with:
\begin{eqnarray}
\label{fluctua}
\Pi(x,y)&=&\Pi^{\rm ring}(x,y)+\sum_{kl}{\rm Tr}\left(\Gamma_k\Gamma_l\right)
\int du\,dv\, G^{\rm ring}_{\pi}(u-v) \Pi^0(x,u,y,v) \nonumber \\
& & +\sum_{kl}{\rm Tr}\left(\Gamma_k\Gamma_l\right)
\int du\,dv\,G^{\rm ring}_{\pi}(u-v) \left[\Pi^0(x,u,v,y)+\Pi^0(x,y,v,u)\right] \nonumber \\
& &+\int du\,dv\,dw\,ds\, G^{\rm ring}_{\pi}(u-w) G^{\rm ring}_{\pi}(v-s)\Pi^0(x,u,v)
\nonumber \\
& &\times \sum_{klmn}\left[{\rm Tr}\left(\Gamma_k\Gamma_l\Gamma_m\Gamma_n\right)
\Pi^0(y,w,s)+{\rm Tr}\left(\Gamma_k\Gamma_l\Gamma_n\Gamma_m\right)
\Pi^0(y,s,w)\right] .
\end{eqnarray}

We remind that the second derivative, with respect to the external source, of 
$S^B_{\rm eff}\left[\bbox\phi^{\rm ring}\right]$ and 
$S^B_{\rm eff}\left[\bbox\phi^0\right]$, with $\bbox\phi^{\rm ring}$
$\left[\bbox\phi^0\right]$ given by eq.~(\ref{soluzring}) [eqs.~(\ref{soluz}),
(\ref{AA}), (\ref{BB})], gives the same result 
(the ring polarization propagator) when evaluated ad 
$\bbox\varphi=0$.

\begin{figure}
\begin{center}
\mbox{\epsfig{file=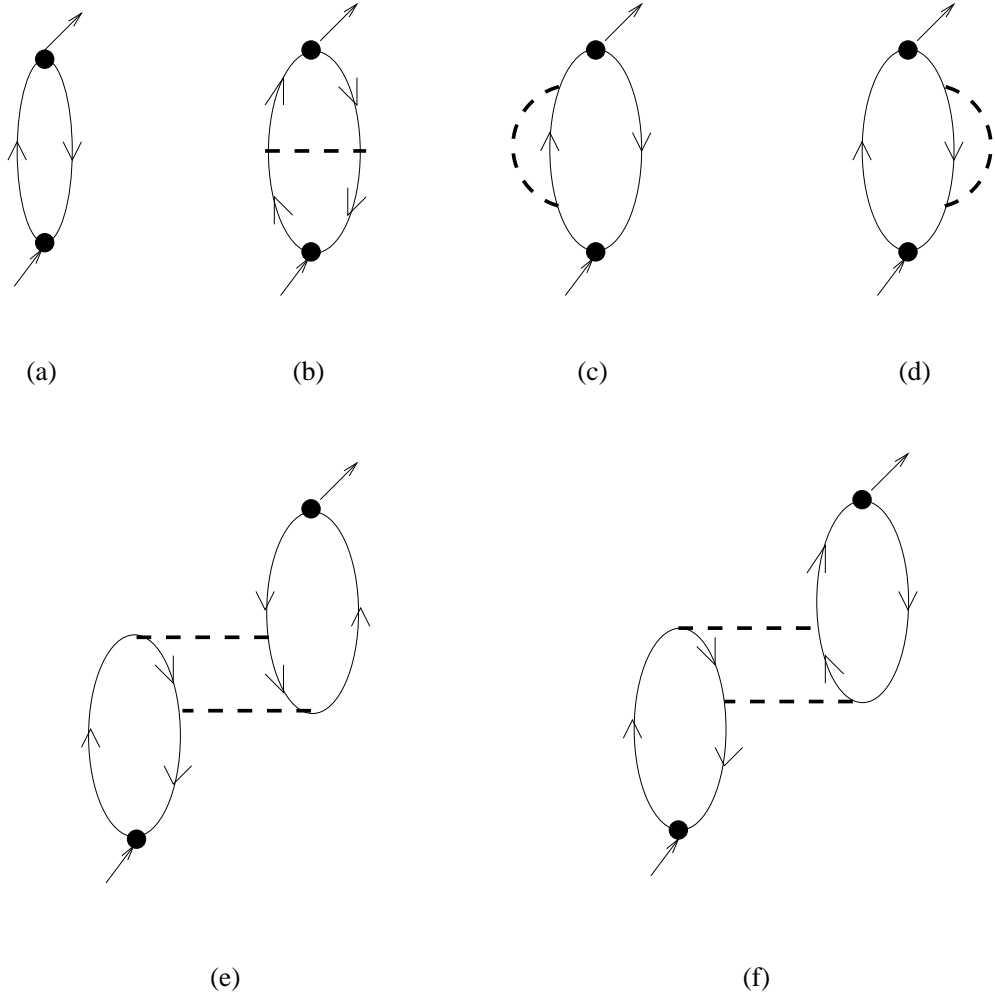,width=.8\textwidth}}
\vskip 5mm
\caption{Feynman diagrams for the polarization propagator of 
eq.~\protect(\ref{fluctua}): (a) particle-hole;
(b) exchange; (c) and (d) self-energy-type; (e) and (f) correlation diagrams. 
Only the first contribution to the {\sl p-h} ring expansion has been drawn. The
dashed line represent ring-dressed pion propagators.}
\label{oneloop}
\end{center}
\end{figure}

\begin{figure}
\begin{center}
\mbox{\epsfig{file=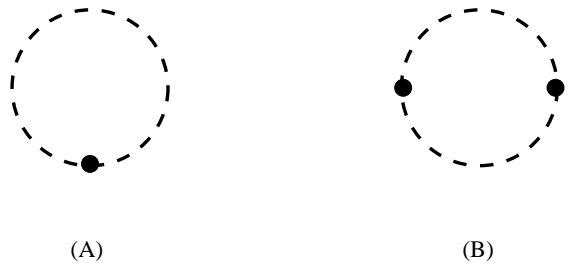,width=.45\textwidth}}
\vskip 5mm
\caption{First order diagrams in the bosonic loop expansion. Diagrams 
(b), (c), (d) of fig.~\protect\ref{oneloop} reduce to diagram (A), 
while (e), (f) to (B).}
\label{oneloop1}
\end{center}
\end{figure}

The Feynman diagrams corresponding to eq.~(\ref{fluctua}) are 
depicted in fig.~\ref{oneloop}.
Diagram (a) represents the Lindhard function $\Pi^0(x,y)$, 
which is the first term of $\Pi^{\rm ring}(x,y)$. In (b) we have an 
exchange diagram (the dashed lines represent ring-dressed
pion propagators); (c) and (d) are self-energy diagrams, while in (e) and (f)
we show the correlation diagrams of our approximation. The approximation
scheme we have developed is also referred to as bosonic loop expansion
(BLE). The practical
rule to classify the Feynman diagrams according to their order in the
BLE is to reduce to a point all its fermionic lines and
to count the number of bosonic loops left out. In our case the diagrams 
(b)-(f) of fig.~\ref{oneloop} reduce to a one-boson-loop.
Diagrams (b), (c), (d) can be represented by the loop (A) of 
fig.~\ref{oneloop1}, while (e), (f) correspond to the loop (B).

When we include in the model the excitation of barionic resonances,
then we have to replace the fermionic field with multiplets. The topology
of the diagrams remains the same as in fig.\ref{oneloop} but, introducing
for example the $\Delta$ resonance,
each fermionic line stands for a nucleon or for a $\Delta$, taking care of
isospin conservation. One thus obtain 15 exchange, 30 self-energy
and 98 correlation diagrams (see ref.~\cite{Ce97} for the whole diagrammology).

Moreover, since this expansion is characterized by the topology
of the diagrams, we can include in the model additional mesonic degrees of
freedom together with phenomenological short range correlations.
In particular, the extension to other spin-isospin channels simply amounts to change
the definition of the $\Gamma_i$ and the same occurs for the non-relativistic 
reduction of the theory. Accordingly, for the non-relativistic 
pion exchange $\Gamma_i$
becomes $(\bbox \sigma\cdot {\bbox q})\tau_i$, for the $\rho$ 
exchange it reads $(\bbox \sigma\times {\bbox q})_k \tau_i$, $k$ being a 
spatial index, and for the $\omega$ exchange 
$\Gamma_i=(\bbox \sigma\times {\bbox q})_i$. The exchange of $\omega$-mesons
is taken into account only inside the one-boson-loop diagrams (b)-(f) of fig.~\ref{oneloop},
but not in the mesonic lines stemming from the $\Lambda$ decay vertex, where the exchanged
meson has to be an isovector ($\pi$ and $\rho$). Beyond $\pi$, $\rho$ and $\omega$ mesons,
the present approach also contains (partly) the exchange of the scalar-isoscalar $\sigma$ meson:
as in the language of the Bonn potential \cite{Ma87}, the latter is described 
through box diagrams (contained
in the correlation diagrams of fig.~\ref{oneloop}), namely by the exchange of
two pions with the simultaneous excitation of one or two of the intermediate
nucleons to a $\Delta$.

A further difficulty arises if we start from a potential theory, instead of one 
containing bosons as true degrees of freedom. This disease is however easily
overcome by means of a Hubbard-Stratonovitch transformation, 
which enables us to substitute
a potential with an interaction of the nucleons with a suitably introduced 
auxiliary field. For a scalar-isoscalar potential $V$ the relevant identity reads:
\begin{eqnarray}
\label{HSI}
&\exp\left\{\displaystyle \frac{i}{2}\int dx\,dy\,\overline\psi(x)
\psi(x)V(x-y)\overline\psi(y)\psi(y)\right\}=& \nonumber \\
&\sqrt{\rm det\, V}\displaystyle \int {\cal D}[\sigma]\exp\left\{\frac{i}{2}
\int dx\,dy\,\sigma(x)V^{-1}(x-y)
\sigma(y)+i\int dx\, \overline\psi(x)\psi(x)\sigma(x)\right\} ,
\end{eqnarray}
where $\sigma$ is the auxiliary field.
Clearly, the previous derivation will remain valid, providing one substitutes
the inverse propagator of the auxiliary field with the inverse potential in
the ``free'' part of the action.

Finally, a relevant point for the feasibility of the calculations is that
all fermion loops in fig.~\ref{oneloop} can be evaluated analytically \cite{Ce88,Ce92},
so that each diagram reduces to a 3-dimensional (numerical) integral.

In particular, the formalism can be applied to evaluate
both the longitudinal and the transverse functions $U_{L,T}$ of 
eq.~(\ref{propU}), which are required in eqs.~(\ref{Sigma2}), (\ref{Alpha}). In the
approximation of fig.~\ref{oneloop} we have to replace eq.~(\ref{Alpha}) with: 
\begin{eqnarray}
\label{Alpha1}
{\alpha}(q)&=&\left\{S^2+\frac{P^2}{m_{\pi}^2}\bbox q^2\right\}F_{\pi}^2(q)
G_{\pi}^0(q)+\frac{\tilde{S}^2(q)U_1(q)}{1-V_L(q)U_1(q)} \nonumber \\
& & +\frac{\tilde{P}_L^2(q)U_1(q)}{1-V_L(q)U_1(q)}+
\frac{\tilde{P}_T^2(q)U_1(q)}{1-V_T(q)U_1(q)} \nonumber \\
& & +\left[\tilde S^2(q)+\tilde P_L^2(q)\right]U_{2,L}(q)+\tilde P_T^2(q)
U_{2,T}(q) ,
\end{eqnarray}
where 
\beq
U_1=U^{ph}+U^{\Delta h} ,
\eeq
and $U_{2,L}$, $U_{2,T}$ are evaluated from diagrams (b)-(f) of 
fig.~\ref{oneloop} (taking into account the $\Delta$
excitation as well) using the standard Feynam rules. The normalization of
these functions is such that $U^{ph}(x,y)=4\Pi^0(x,y)$, with $\Pi^0$
given by eq.~(\ref{pizero}).

\section{Results and discussion}
\label{res}
In the present work the evaluation of the hypernuclear decays is 
performed in nuclear matter. However, in order to compare with experimental
data on finite (real) nuclei, we employ different Fermi momenta, 
which are fixed on the following basis. In local density approximation the ``local''
Fermi momentum $k^A_F(r)$ is related to the nuclear density:
\beq
{\rho}_A(r)=\displaystyle \frac {{\rho}_0(A)}{1+e^{[r-R(A)]/a}} , \hspace {0.3 in}
{\rho}_0(A)=\displaystyle \frac {A}{\frac{4}{3} \pi R^3(A)
\{1+[\frac{\pi a}{R(A)}]^2\}}
\vspace{0.3 in}
\eeq
($a=0.54$ fm, $R(A)=1.12A^{1/3}-0.86A^{-1/3}$ fm), by the equation:
\beq
\label{local}
k^A_F(r)=\left[\frac{3}{2}\pi^2 \rho_A(r)\right]^{1/3} .
\eeq 
For the present purpose, an average, {\it fixed} Fermi momentum can be
obtained by weighting each local $k_F$ with the density itself:
\beq
\label{kf1}
<k_F>_A=\frac{1}{A}\int d\bbox r k^A_F(r)\rho_A(r) .
\eeq
Alternatively (and more realistically) the average Fermi momentum should be 
determined by the probability density of the $\Lambda$ inside the nucleus, 
according to the following definition:
\beq
\label{kf3}
<k_F>_A=\int d\bbox r k^A_F(r)|\psi_{\Lambda}(\bbox r)|^2 ,
\eeq 
$\psi_{\Lambda}$ being the $\Lambda$ wave function. We have calculated
the latter from a Wood-Saxon
potential with thickness $a=0.6$ fm and radius and depth which
reproduce the measured $s$ and $p$ $\Lambda$-levels \cite{Al99}.
Since the $\Lambda$ wave function is preferably located in the interior
of the nucleus (in fact, the hyperon occupies the $1s$ level), we expect
larger $<k_F>$ values from the prescription (\ref{kf3}) than from (\ref{kf1}).

We classify the hypernuclei for which are available
experimental data on the non-mesonic decay rate into three mass regions
(medium-light: $A\simeq 10$; medium: $A\simeq 30\div 60$;
and heavy hypernuclei: $A\simeq 200$),
as shown in table~\ref{kmed}.
The experimental bands include values of the non-mesonic widths 
which are compatible with all the quoted experiments.
In the third and fourth columns we report the corresponding $<k_F>$, as calculated
with eqs.~(\ref{kf1}) and (\ref{kf3}), respectively.
\begin{table}[t]
\begin{center}
\caption{Average Fermi momenta.}
\vspace{0.3cm}
\label{kmed}
\begin{tabular}{c|c c c}
\mc {1}{c|}{} &
\mc {1}{c}{$\Gamma^{\rm exp}_{NM}$} &
\mc {1}{c}{$<k_F>$} eq.~(\ref{kf1})&
\mc {1}{c}{$<k_F>$} eq.~(\ref{kf3}) \\
\mc {1}{c|}{} &
\mc {1}{c}{} &
\mc {1}{c}{(fm$^{-1}$)} &
\mc {1}{c}{(fm$^{-1}$)} \\ \hline\hline
Medium-Light: $^{11}_{\Lambda}$B - $^{12}_{\Lambda}$C & $0.94\div 1.07$ \cite{Sz91,No95}& 0.95 &  1.08 \\
Medium : $^{28}_{\Lambda}$Si - $^{56}_{\Lambda}$Fe    & $1.20\div 1.30$ \cite{Bh98}& $\simeq 1.1$ &  $\simeq 1.2$\\
Heavy: $^{209}_{\Lambda}$Bi - $^{238}_{\Lambda}$U     & $1.45\div 1.70$ \cite{Ku98,Ar93,Oh98}& 1.21 &  1.36\\
\end{tabular}
\end{center}
\end{table}
Accordingly, in the calculations of the decay width we use
$k_F=1.1$ fm$^{-1}$ for medium-light, $k_F=1.2$ fm$^{-1}$ for medium and 
$k_F=1.36$ fm$^{-1}$ for heavy hypernuclei.

Before discussing the results, we list here the parameters which
enter the calculation of the hypernuclear width. We point out that their 
values (with the exception of the Landau parameters $g^{\prime}$ and 
$g^{\prime}_{\Lambda}$, which will be discussed separately) have not been
taken as free parameters. Instead, we kept them fixed on the basis of the
existing phenomenology (for example in the analysis of the quasi-elastic
electron-nucleus scattering, of the spin-isospin nuclear response functions,
etc).
\begin{itemize}
\item The correlation momentum is set to $q_c=0.78$ GeV;
\item The cut-offs in the pion, $\rho$ and $\omega$ form factors are:
$\Lambda_{\pi NN}=1.3$ GeV, 
$\Lambda_{\pi N\Delta}=\Lambda_{\pi \Delta \Delta}=1$~GeV,
$\Lambda_{\rho NN}=\Lambda_{\rho N\Delta}=\Lambda_{\rho \Delta \Delta}=2.5$ GeV,
$\Lambda_{\omega NN}=\Lambda_{\omega \Delta \Delta}=1$~GeV;
\item The pion coupling constants are: $f^2_{\pi NN}/4\pi=0.08$, 
$f^2_{\pi N\Delta}/4\pi=0.32$, $f^2_{\pi \Delta \Delta}/4\pi=0.016$;
\item The rho coupling constants are (see eq.~(\ref{rhocoup})): 
$C_{\rho NN}=C_{\rho N \Delta}=C_{\rho \Delta \Delta}=2.3$;
\item The $\omega$ coupling constants are: 
$C_{\omega NN}=C_{\omega \Delta \Delta}=1.5$ (renormalized values);
\item Finally, the difference between the nucleon and $\Lambda$ binding
energies, which enter eq.~(\ref{Sigma2}) as $\Delta V=V_{\Lambda}-V_N$,
has been fixed, for the heaviest nuclei ($k_F=1.36$ fm$^{-1}$), from the
corresponding depths of the binding potential
($V_N=-55$ MeV, $V_{\Lambda}=-32$ MeV). Then, by assuming that 
$\Delta V$ is roughly proportional to the density, we obtain
$\Delta V=18$ MeV at $k_F=1.2$ fm $^{-1}$ and $\Delta V=12$ MeV
at $k_F=1.1$ fm$^{-1}$. It is worth pointing out, however, that the 
resulting widths are only weakly affected by the precise value of $\Delta V$
(a change of $5$ MeV in $\Delta V$ would result in a 3\% variation in the
decay rates).
\end{itemize}

The numerical evaluation of the first order contributions in the BLE
[diagrams (b)-(f) in fig.~\ref{oneloop}] presents a serious problem
when the mesons ($\pi$, $\rho$ and $\omega$ in our case) are dynamic. 
The pionic branch in RPA is 
coupled to the particle-hole mode. By using non-relativistic kinematics for
the nucleons, the relativistic pionic branch enters
the particle-hole continuum for a momentum of about $1.6$ GeV. This fact
generates large oscillations in the calculation of the one-loop diagrams (where the
momentum of the mesons is integrated out), 
which are absolutely unphysical and unavoidable
from the numerical point of view. To prevent this mishap, 
we are forced to use static mesons
in the evaluation of the one-boson-loop contributions. In all the
mesonic lines of fig.~\ref{self11} which are ``external'' to the bosonic loop
we can use, on the contrary, 
the correct dynamical prescription, the relevant momenta being limited to
about $0.6$ GeV.

An important ingredient in the calculation of the $\Lambda$ decay rates
is the short range part of the $NN$ and $\Lambda N$ interactions: in fact, in
the non-mesonic processes the exchanged momentum
is very large (about $400$ MeV). We parametrize these short range correlation
with the functions reported in the appendix of ref.~\cite{Al99}. The
zero energy and momentum limits of these correlations, $g^{\prime}$ and
$g^{\prime}_{\Lambda}$ (which we call, for hystorical reasons, Landau parameters), are considered
as {\it free parameters}. No experimental constraints 
are available on $g^{\prime}_{\Lambda}$, while 
in the framework of the ring approximation (namely by neglecting the 
{\sl 2p-2h} states in the $\Lambda$ self-energy), 
realistic values of $g^{\prime}$ lie in the
range $0.6\div 0.7$ \cite{Os82}. However, in the present context $g^{\prime}$
correlates not only {\sl p-h} pairs, but also {\sl p-h} with {\sl 2p-2h} 
states and
besides, in some diagrams [for instance (d) and (e) of fig.~\ref{self11}], 
two consecutive
$g^{\prime}$ are connected to the same fermionic line, introducing some double
counting, namely a renormalization of $g^{\prime}$. In the picture of 
figures~\ref{self11} and \ref{oneloop} 
the $\Lambda$ self-energy acquires an energy and momentum 
behaviour which cannot be explained and simulated on the basis of the
simple ring approximation.
Therefore, the physical meaning of the Landau parameters is different 
in our scheme with respect to the customary phenomenology, hence in the present paper 
we will use $g^{\prime}$ as free parameters, to be fixed in order to reproduce
the experimental hypernuclear decay rates.

As discussed in ref.~\cite{Al99}, for fixed $g^{\prime}$ the
non-mesonic width (the total width in nuclear matter, where $\Gamma_M=0$)
has a minimum as a function of $g^{\prime}_{\Lambda}$, which
is almost indipendent of the value of $g^{\prime}$. This characteristic does not
depend on the set of diagrams taken into account in the calculation, but it is simply
due to the interplay between the longitudinal and transverse parts of the 
$p$-wave $\Lambda N\rightarrow NN$ potential
[$\tilde P_L$ and $\tilde P_T$ functions of eqs.~(\ref{Alpha}), (\ref{Alpha1})]. 
In the present calculation 
the minimum is obtained, again, for $g^{\prime}_{\Lambda}\simeq 0.4$. 

\begin{figure}
\begin{center}
\mbox{\epsfig{file=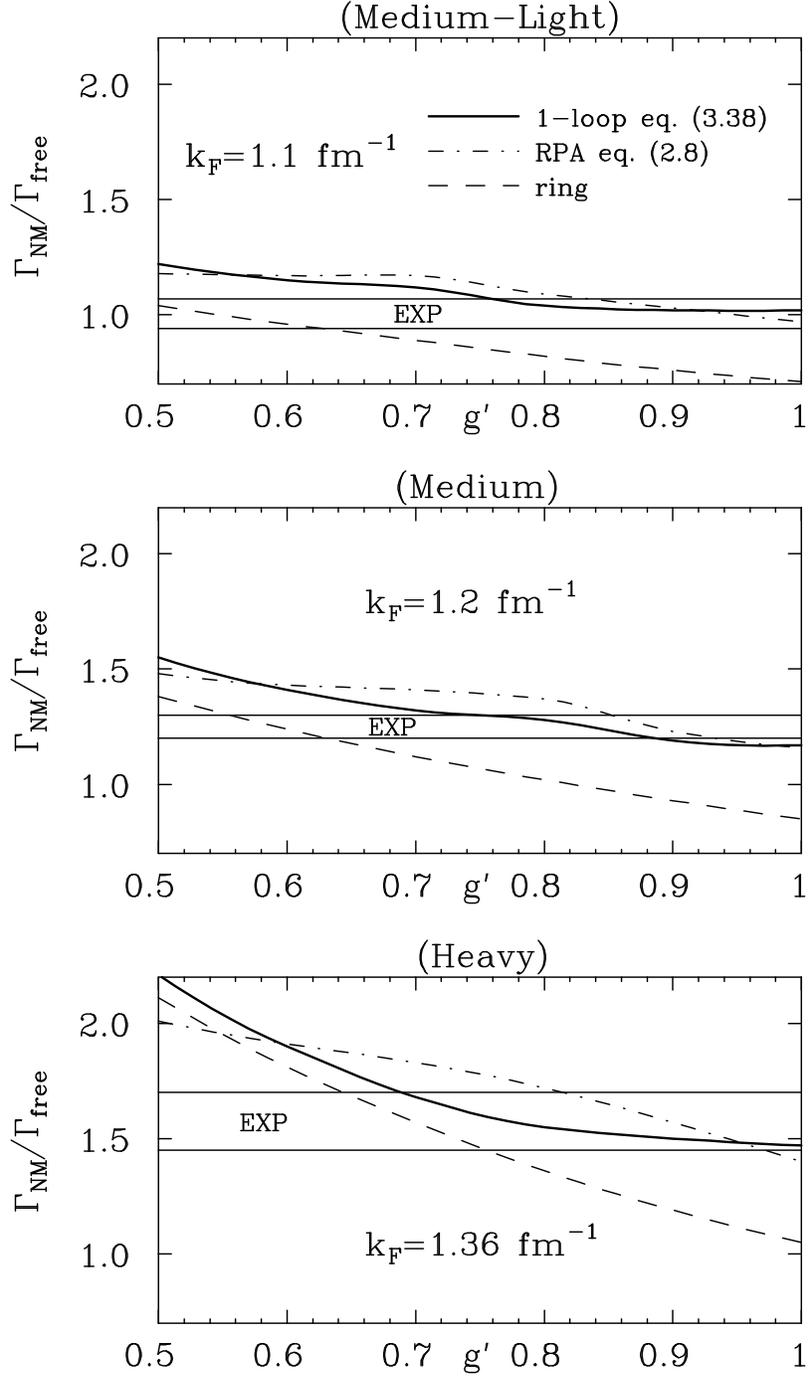,width=0.87\textwidth}}
\vskip 4mm
\caption{Dependence of the non-mesonic width on the Landau 
parameter $g^{\prime}$, for $g^{\prime}_{\Lambda}=0.4$. The three plots 
correspond to the classification of
table~\ref{kmed}. The thick solid curves refer to the
one-boson-loop approximation of eq.~(\ref{Alpha1}), 
the dot-dashed ones to the RPA calculation of eq.~(\ref{Alpha}) and
the dashed ones to the ring approximation.
The experimental bands of table~\ref{kmed} lies in between the horizontal 
solid lines.}
\label{gammagprimo}
\end{center}
\end{figure}

\begin{figure}
\begin{center}
\mbox{\epsfig{file=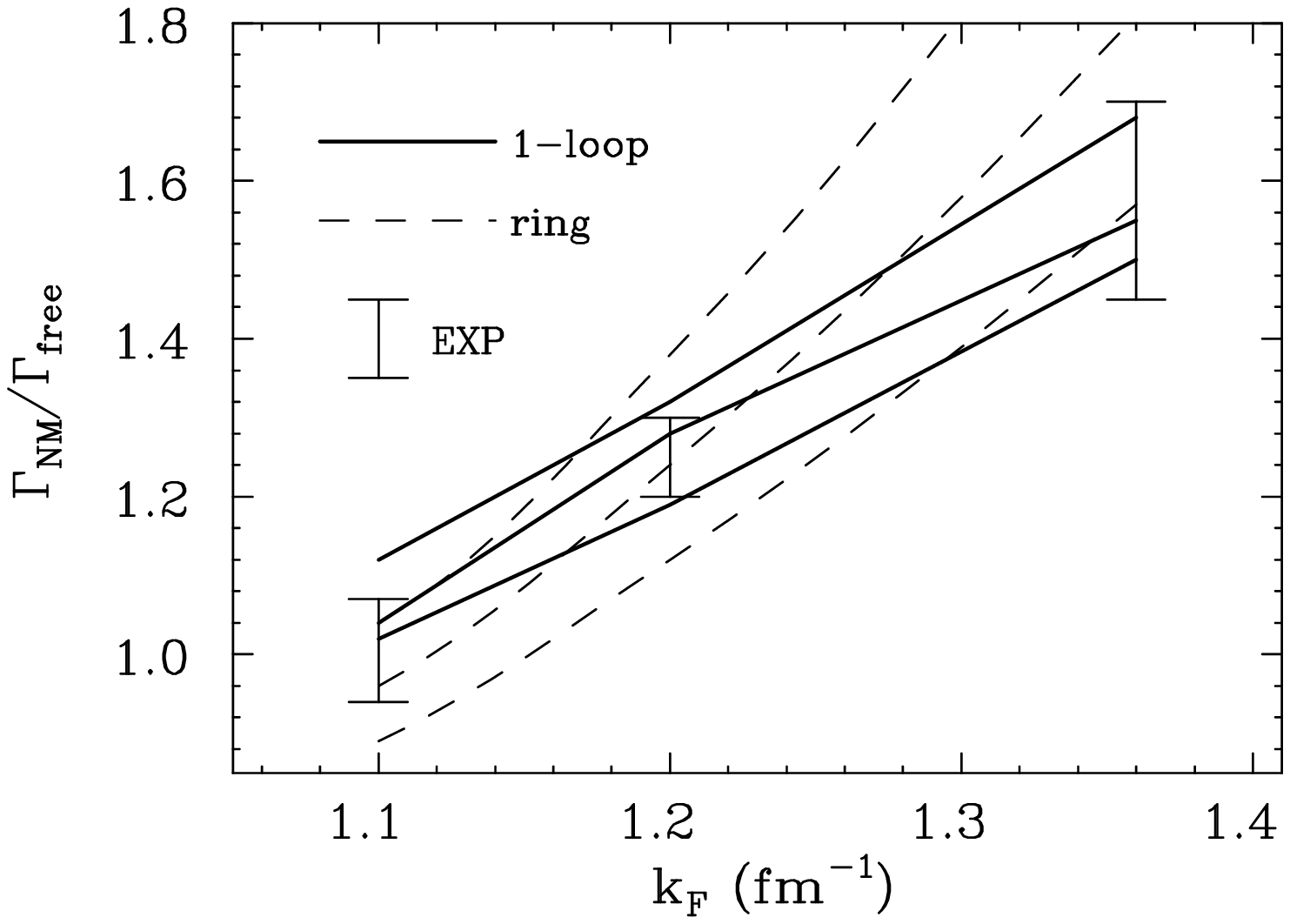,width=0.7\textwidth}}
\vskip 2mm
\caption{Dependence of the non-mesonic width on the Fermi momentum.
The solid curves refer to the one-boson-loop approximation
(with $g^{\prime}=0.7, 0.8, 0.9$ from the top to the bottom), 
while the dashed lines refer to the ring approximation ($g^{\prime}=0.5, 0.6, 0.7$). 
The experimental data are also shown.}
\label{gammakf}
\end{center}
\end{figure}

Fixing $g^{\prime}_{\Lambda}=0.4$, in ring approximation
we can reproduce the experimental decay rates
by using $g^{\prime}$ values which are compatible with the existing literature. 
In figure~\ref{gammagprimo} we show, as a function of $g^{\prime}$
(for $g^{\prime}_{\Lambda}=0.4$), the calculated non-mesonic decay widths
(in unit of the free $\Lambda$ width) 
for the three mass regions of table~\ref{kmed}.
The thick solid curves refer to the one-boson-loop approximation
of eq.~(\ref{Alpha1}) and fig.~\ref{oneloop}, while the dot-dashed curves are 
obtained through an RPA iteration of both the particle-hole and the 
one-boson-loop diagrams, namely by using eq.~(\ref{Alpha}). 
However, we remind that only the former approximation
has a theoretically founded basis, in line with the scheme 
introduced in sec.~\ref{func}: indeed, the present RPA calculation has the
tendency to overestimate, in the acceptable range of $g^{\prime}$ values, 
the experimental non-mesonic widths. The dashed lines represent the pure
ring approximation. The calculation is compatible with the experimental bands
for the $g^{\prime}$ values reported in table~\ref{gprimo}. 
\begin{table}[t]
\begin{center}
\caption{$g^{\prime}$ values compatible with the experiments.}
\vspace{0.3cm}
\label{gprimo}
\begin{tabular}{c|c c}
\mc {1}{c|}{} &
\mc {1}{c}{1-loop} &
\mc {1}{c}{ring} \\ \hline\hline
$k_F=1.1$ fm$^{-1}$  & $\gtrsim 0.75$     & $0.45\div 0.65$ \\
$k_F=1.2$ fm$^{-1}$  & $0.75\div 0.90$ & $0.55\div 0.65$ \\
$k_F=1.36$ fm$^{-1}$ & $0.70\div 1.00$ & $0.65\div 0.75$ \\
\end{tabular}
\end{center}
\end{table}
As we have already noticed, the intervals corresponding 
to the ring calculation are in agreement
with the phenomenology of other processes, like the $(e,e')$ quasi-elastic
scattering. 
However, only the full calculation (column 1-loop) allows for
a good description (with the same $g^{\prime}$ values) 
of the rates in the whole range of $k_F$ considered here. In fig.~\ref{gammakf}
we see the dependence of the non-mesonic widths on the Fermi momentum.
The solid lines correspond to the one-loop approximation, with 
$g^{\prime}=0.7, 0.8, 0.9$ from the top to the bottom, while the dashed
lines refer to the ring approximation, with $g^{\prime}=0.5, 0.6, 0.7$ 
(again from the top
to the bottom). We can then conclude that for the one-loop calculation 
the best choice of the Landau parameters is the following:
\beq
g^{\prime}=0.8, \hspace{0.6cm} g^{\prime}_{\Lambda}=0.4 .
\eeq
This parametrization is the same which was employed in a different
theoretical framework \cite{Al99} to reproduce the
experimental decay rates in the range 
$^{12}_{\Lambda}$C - $^{238}_{\Lambda}$Pb.
However, we must point out that in the previous calculation the
{\sl 2p-2h} contributions in the $\Lambda$ self-energy were evaluated
by using a phenomenological parametrization of the pion-nucleus
optical potential, also accounting for the available phase space for the
{\sl 2p-2h} states. The role played by
the Landau parameters is different in the
two approaches. In the present paper we have microscopically evaluated
all the relevant diagrams which contribute at the one-boson-loop level;
however, due to the very long computing times, we cannot 
implement the calculation for finite nuclei through the local density 
approximation used in \cite{Al99}.

\begin{figure}
\begin{center}
\mbox{\epsfig{file=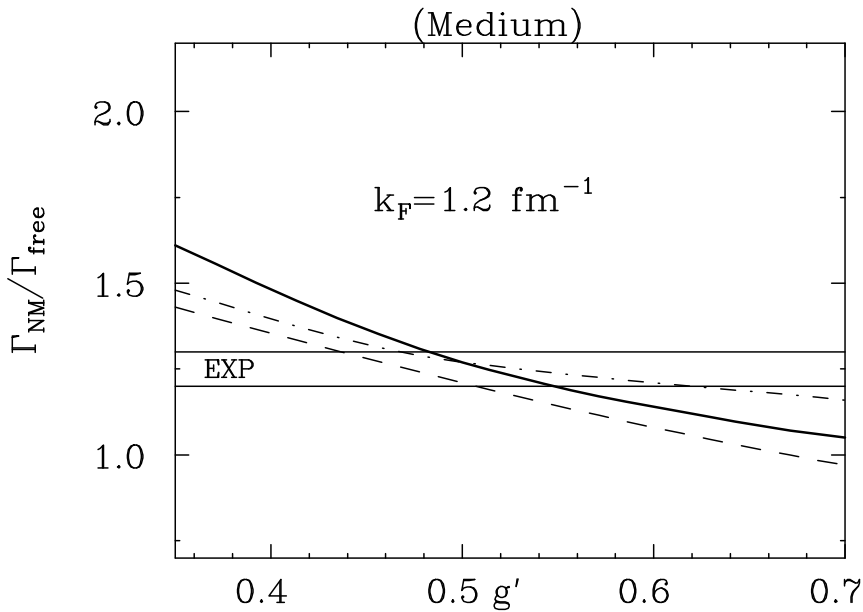,width=0.70\textwidth}}
\caption{Dependence of the non-mesonic width on the Landau 
parameter $g^{\prime}$, for $g^{\prime}_{\Lambda}=0.4$ and 
$k_F=1.2$ fm$^{-1}$ in the static limit. The various curves have the same
meaning as in fig.~\protect\ref{gammagprimo}.}
\label{static}
\end{center}
\end{figure}

The fairly large value of $g^{\prime}$, which we adopt in order to 
reproduce the $\Lambda$ widths in the one-boson-loop approximation,
deserves some comment. Indeed, beside the above mentioned phenomenology
within the framework of ring approximation, which favoured values of 
$g^{\prime}$ between 0.6 and 0.7, previous calculations of the $(e,e^{\prime})$
inclusive longitudinal response function within the one-boson-loop approximation
\cite{Ce97} employed a rather small value ($g^{\prime}=0.35$) 
of this parameter. 
We remind that the longitudinal response only concerns the scalar-isoscalar
and scalar-isovector channels: hence neither the $\pi$ or $\rho$ mesons
nor $g^{\prime}$ enter the particle-hole interaction lines external
to the one bosonic loop diagrams. Therefore, in ref.~\cite{Ce97} $g^{\prime}$
was employed {\it only} in conjunction with static mesons. 
In order to understand the influence in the present calculation
of the dynamical (external) pion propagation, we show in figure~\ref{static}
the results of a completely static calculation
(also the ``external'' mesonic lines are static). The non-mesonic widths we
obtain in this case are smaller than in figure \ref{gammagprimo} and the 
experimental band is generally reproduced by using smaller $g^{\prime}$
values: tipically $g^{\prime}=0.5$ in the one-loop approximation.
This outcome somehow reconciles the parametrization of the short range correlations
used in the evaluation of 
the hypernuclear decay rates with the one of ref.~\cite{Ce97} 
for the nuclear responses in the inclusive electron scattering.

In table~\ref{newold} we show the comparison between the
one-boson-loop approximation (OBL) and the phenomenological model (PM) of 
ref.~\cite{Al99} at fixed $k_F$. The calculations have been carried out 
in both cases with $g^{\prime}=0.8$ and
$g^{\prime}_{\Lambda}=0.4$. For technical reasons, the OBL calculation does not 
allows to precisely identify
the partial ratios $\Gamma_1$ and $\Gamma_2$ which contribute to the total 
$\Gamma_{NM}=\Gamma_1+\Gamma_2$. In fact, we cannot separate in the imaginary parts of 
the diagrams (b)-(f) of fig.~\ref{oneloop} the contributions coming
from cuts on {\sl p-h} and {\sl 2p-2h} states, and hence the partial width 
($\Gamma_2$) stemming from the two-nucleon induced decay.
The values listed in the 
table for $\Gamma^{\rm OBL}_2$ have been obtained 
from the total imaginary part of the diagrams \ref{oneloop}(b)-\ref{oneloop}(f) 
[namely from the last two terms in the r.h.s.~of eq.~(\ref{Alpha1})]. 
In this approximation,
$\Gamma^{\rm OBL}_1=\Gamma_{\rm ring}$ [second, third and fourth 
terms in the r.h.s.~of eq.~(\ref{Alpha1})]. 
As a matter of fact, one would expect that $\Gamma_2$
increases with $k_F$ (and this is the case for the PM calculation), 
but the OBL results
do not follow this statement. From table~\ref{newold} and from the study of the
$g^{\prime}$-dependence of $\Gamma^{\rm OBL}_2$,
the only conclusion we can draw on the two-body induced process
in OBL approximation is that for 
$1.1$~fm$^{-1}\lesssim k_F\lesssim 1.36$~fm$^{-1}$ and $0.5\lesssim g^{\prime}\lesssim 1$, 
$\Gamma_2/\Gamma_{\rm free}=0.1\div 0.3$.
\begin{table}[t]
\begin{center}
\caption{Comparison between the one-boson-loop approximation (OBL) and the 
phenomenological model (PM) of ref.~\protect\cite{Al99} for $g^{\prime}=0.8$,
$g^{\prime}_{\Lambda}=0.4$. The decay rates are in unit of the free $\Lambda$ width.}
\vspace{0.3cm}
\label{newold}
\begin{tabular}{c|c c| c c| c c}
\mc {1}{c|}{} &
\mc {2}{c|}{$k_F=1.1$ fm$^{-1}$} &
\mc {2}{c|}{$k_F=1.2$ fm$^{-1}$} &
\mc {2}{c}{$k_F=1.36$ fm$^{-1}$} \\
\mc {1}{c|}{} &
\mc {1}{c}{OBL} &
\mc {1}{c|}{PM} &
\mc {1}{c}{OBL} &
\mc {1}{c|}{PM} &
\mc {1}{c}{OBL} &
\mc {1}{c}{PM} \\ \hline\hline
 $\Gamma_1$          & 0.82            & 0.81 & 1.02 & 1.00 & 1.36 & 1.33 \\
 $\Gamma_2$          & 0.22            & 0.13 & 0.26 & 0.18 & 0.19 & 0.26 \\
 $\Gamma_{NM}$       & 1.04            & 0.94 & 1.28 & 1.19 & 1.55 & 1.59 \\
\end{tabular}
\end{center}
\end{table}

\section{Conclusions}
\label{concl}
The $\Lambda$ non-mesonic decay widths have been evaluated in nuclear matter
within the polarization propagator method. The Feynman diagrams contributing to 
the $\Lambda$ self-energy have been classified,
via a functional approach, by the first order approximation 
of the bosonic loop expansion. In this scheme,
the two-body induced decay $\Lambda NN\rightarrow NNN$ is taken into account
for the first time from a fully microscopic point of view.

In order to compare the calculation with the available 
experimental data, we have employed
different ``average'' Fermi momenta 
for the following three mass regions:
medium-light ($A\simeq 10$); medium ($A\simeq 30\div 60$);
and heavy hypernuclei ($A\simeq 200$). 

The only free parameters of our model are $g^{\prime}$ and $g^{\prime}_{\Lambda}$,
which incorporate the $NN$ and $\Lambda N$ short range interactions, respectively.
We have fixed these Laudau parameters in order to reproduce the observed decay 
widths, and, in the one-boson-loop approximation,
the best choice has turned out to be $g^{\prime}=0.8$, 
$g^{\prime}_{\Lambda}=0.4$.
The agreement between the experimental widths and the theoretical evaluation
is of the same quality as in ref.~\cite{Al99}; interestingly, the same values of
$g^{\prime}$ and $g^{\prime}_{\Lambda}$ give the best fit to the data in 
both approaches. However, we point out that the role of these Landau parameters
is slightly different in the two cases. 
One difference, with respect to ref.~\cite{Al99},
in the role played by $g^{\prime}$ and $g^{\prime}_{\Lambda}$ can be
understood by comparing equations (\ref{Alpha}) and (\ref{Alpha1}):
clearly, in the former (employed in ref.~\cite{Al99}) 
the short range correlations have a major influence
in renormalizing the full polarization propagator $U_{L,T}$, while in the latter
(used in the present calculation) the RPA renormalization 
only affects the Lindhard function $U_1$. 
Moreover, in the present calculation $g^{\prime}$
comes about together with static mesons 
(because of the discussed computational problems) 
in the lines ``internal'' to the bosonic loops and 
together with dynamical mesons in the lines
``external'' to the bosonic loops, while in \cite{Al99} all mesonic lines are 
dynamical. We also note that,
by introducing a completely static description,
the experimental data are reproduced with smaller $g^{\prime}$ values
($\simeq 0.5$). This finding is in fairly good agreement with 
the parametrization introduced in ref.~\cite{Ce97} for the BLE
calculation of the nuclear response functions in the inclusive electron scattering,
where $g^{\prime}$ only correlates static mesons, as explained
in the previous section.

Finally, while the estimated $\Gamma^{\rm OBL}_1$ only contains one-body
induced decays (in ring approximation), contributions to the total width stemming
from this channel are also embodied in $\Gamma^{\rm OBL}_2$, but cannot be explicitely 
separated from the truly two-body induced ones. On the basis of reasonable
extimates, the two-body stimulated decay is found to be sizeable,
($\Gamma_2/\Gamma_{\rm free}=0.1\div 0.3$ for the whole hypernuclear mass spectrum
studied), with values comparable with the results of ref.~\cite{Al99}.

In conclusion, the present calculation satisfactorily reproduces
the non-mesonic decay widths for $A\gtrsim 10$ within a fully microscopic 
approach at fixed, average density, taking into account both one-body and
two-body induced processes.

\acknowledgments

We acknowledge financial support from the MURST.

\vfill\eject


\newpage
\begin{thebibliography}{99}
\bibitem{Al99} W. M. Alberico, A. De Pace, G. Garbarino and A. Ramos, 
{\bf nucl-th/9902023}.
\bibitem{Os93} E. Oset and J. Nieves, {\em Phys. Rev.} {\bf C47} (1993) 1478.
\bibitem{It95} K. Itonaga, T. Ueda and T. Motoba,
{\em Nucl. Phys.} {\bf A585} (1995) 331c.
\bibitem{Ne82} J. W. Negele, {\em Rev. Mod. Phys.} {\bf 54} (1982) 813.
\bibitem{Al87} W. M. Alberico, R. Cenni, A. Molinari and P. Saracco,
{\em Ann. Phys.} {\bf 174} (1987) 131.
\bibitem{Ce97} R. Cenni, F. Conte and P. Saracco, 
{\em Nucl. Phys.} {\bf A623} (1997) 391.
\bibitem{Os85} E. Oset and L. L. Salcedo, {\em Nucl. Phys.} {\bf A443} (1985) 704.
\bibitem{Ra95} A. Ramos, E. Oset and L. L. Salcedo, 
{\em Phys. Rev.} {\bf C50} (1995) 2314.
\bibitem{Pa98} A. Parre\~no, A. Ramos, C. Bennhold and K. Maltman, {\em Phys. Lett.}
{\bf B 435} (1998) 1.
\bibitem{Pa97} A. Parre\~{n}o, A. Ramos and C. Bennhold,
{\em Phys. Rev.} {\bf C56} (1997) 339.
\bibitem{Du96} J. F. Dubach, G. B. Feldman, B. R. Holstein and L. de la Torre,
{\em Ann. Phys.} {\bf 249} (1996) 146.
\bibitem{Wa71} A. L. Fetter and J. D. Walecka, {\em Quantum Theory of Many
Particle Systems} (McGraw-Hill, New York, 1971).
\bibitem{Os90} E. Oset, P. Fern\'andez de C\'ordoba, L. L. Salcedo and
R. Brockmann, {\em Phys. Rep.} {\bf 188} (1990) 79.
\bibitem{Ma87} R.Machleidt, K.Holinde, and Ch. Elster, 
  {\em Phys. Rep.} {\bf 149} (1987) 1.
\bibitem{Ce88}R. Cenni and P. Saracco, {\em Nucl. Phys.} {\bf A487} (1988) 279.
\bibitem{Ce92}R. Cenni, F. Conte, A. Cornacchia and P. Saracco,
{\em Rivista del Nuovo Cimento} vol. 15 n. 12 (1992).
\bibitem{Sz91} J. J. Szymanski {\sl et al.}, {\em Phys. Rev.} {\bf C43} (1991) 849.
\bibitem{No95} H. Noumi {\sl et al.}, {\em Phys. Rev.} {\bf C52} (1995) 2936.
\bibitem{Bh98} H. C. Bhang {\sl et al.}, {\em Nucl. Phys.} {\bf A639} (1998) 269c.
\bibitem{Ku98} P. Kulessa {\sl et al.}, {\em Phys. Lett.} {\bf B427} (1998) 403.
\bibitem{Ar93} T. A. Armstrong {\sl et al.}, {\em Phys. Rev.} {\bf C47} (1993) 1957.
\bibitem{Oh98} H. Ohm {\sl et al.}, {\em Nucl. Phys.} {\bf A629} (1998) 416c. 
\bibitem{Os82} E. Oset, H. Toki and W. Weise, {\em Phys. Rep.} {\bf 83} (1982) 281.
\end{thebibliography}
\end{document}